\documentclass[aps,prc,twocolumn,nofootinbib,amsmath,showpacs,superscriptaddress,floatfix]{revtex4-1}
\usepackage{graphicx,epsfig,epstopdf,longtable}
\usepackage{mathptmx}
\usepackage[pdftex]{hyperref}
\usepackage{multirow}
\usepackage{wasysym}

\newcommand{\beqy}{\begin{eqnarray}}
\newcommand{\eeqy}{\end{eqnarray}}
\newcommand{\bmlet}{\begin{subequations}}
\newcommand{\emlet}{\end{subequations}}
\newcounter{saveeqn}

\def\gsimeq{\,\,\raise0.14em\hbox{$>$}\kern-0.76em\lower0.28em\hbox  
{$\sim$}\,\,}  
\def\lsimeq{\,\,\raise0.14em\hbox{$<$}\kern-0.76em\lower0.28em\hbox  
{$\sim$}\,\,}  

\begin{document}

\title{First observation of low-energy $\gamma$-ray enhancement in the rare-earth region}

\author{A.~Simon}
	\email{anna.simon@nd.edu}
	\affiliation{Department of Physics, University of Notre Dame, IN 46556-5670, USA}
\author{M.~Guttormsen}
	\email{magne.guttormsen@fys.uio.no}
	\affiliation{Department of Physics, University of Oslo, N-0316 Oslo, Norway}
\author{A.C.~Larsen}
	\email{a.c.larsen@fys.uio.no}
	\affiliation{Department of Physics, University of Oslo, N-0316 Oslo, Norway}
	
\author{C.W.~Beausang}	
	\affiliation{Department of Physics, University of Richmond, Richmond, VA 23171, USA}
\author{P.~Humby}
	\affiliation{Department of Physics, University of Richmond, Richmond, VA 23171, USA}
	\affiliation{Department of Physics, University of Surrey, Surrey, GU27XH, UK}
\author{J.T.~Burke}
	\affiliation{Lawrence Livermore National Laboratory, Livermore, CA 94551, USA}	
\author{R.J.~Casperson}
	\affiliation{Lawrence Livermore National Laboratory, Livermore, CA 94551, USA}		
\author{R.O.~Hughes}
	\affiliation{Lawrence Livermore National Laboratory, Livermore, CA 94551, USA}	
\author{T.J.~Ross} 
	\affiliation{Department of Chemistry, University of Kentucky, Lexington, Kentucky 40506, USA}
	
\author{J.M.~Allmond}
	\affiliation{Physics Division, Oak Ridge National Laboratory, Oak Ridge, TN 37831, USA}
\author{R.~Chyzh}
	\affiliation{Cyclotron Institute, Texas A\&M University, College Station, TX 77843, USA}
\author{M.~Dag}
	\affiliation{Cyclotron Institute, Texas A\&M University, College Station, TX 77843, USA}
\author{J.~Koglin}
	\affiliation{Lawrence Livermore National Laboratory, Livermore, CA 94551, USA}
\author{E.~McCleskey}
	\affiliation{Cyclotron Institute, Texas A\&M University, College Station, TX 77843, USA}
\author{M.~McCleskey}
	\affiliation{Cyclotron Institute, Texas A\&M University, College Station, TX 77843, USA}
\author{S.~Ota}
	\affiliation{Lawrence Livermore National Laboratory, Livermore, CA 94551, USA}
	\affiliation{Department of Physics and Astronomy, Rutgers University, New Brunswick, NJ 08903, USA}
\author{A.~Saastamoinen}
	\affiliation{Cyclotron Institute, Texas A\&M University, College Station, TX 77843, USA}

\date{\today}

\begin{abstract}
The $\gamma$-ray strength function and level density in the quasi-continuum of $^{151,153}$Sm have been measured using BGO shielded Ge clover detectors of the STARLiTeR system. The Compton shields allow for an extraction of the $\gamma$ strength down to unprecedentedly low $\gamma$ energies of $\approx 500$ keV. For the first time an enhanced low-energy $\gamma$-ray strength has been observed in the rare-earth region. In addition, for the first time both the upbend and the well known scissors resonance have been observed simultaneously for the same nucleus. Hauser-Feshbach calculations show that this strength enhancement at low $\gamma$ energies could have an impact of 2-3 orders of magnitude on the (n,$\gamma$) reaction rates for the r-process nucleosynthesis.
\end{abstract}


\maketitle

\section{Introduction}
Atomic nuclei are excellent laboratories for exploring the nature of strongly interacting particles of a finite many-body quantum system.
One of the most useful probes of revealing the nucleus' dynamics is the $\gamma$-ray emission from its excited states. A detailed investigation of $\gamma$-ray transitions at low excitation energies have, for example, shed light on nuclear shape coexistence~\cite{heyde2011}, a pure quantum-mechanical phenomenon without any classical analogue. Furthermore, the emission of high-energy $\gamma$ rays from highly excited nuclei have been a subject of systematic studies throughout the stable isotopes,
with the intriguing discovery that all of them display a Giant Dipole Resonance (GDR), dominated by E1 transitions and centered at $E_\gamma \approx 12-17$ MeV~\cite{dietrich1988}.

In between these two energy regimes, i.e. above the discrete region but below the neutron separation energy $S_n$,
the nuclear dynamics is particularly complex due to the increasing density of states and number of excited quasiparticles. In this excitation-energy region, some rather peculiar $\gamma$-decay patterns have been seen. Close to the $S_n$, an ensemble of states decaying with extraordinarily strong E1 transitions have been found both in stable and exotic, neutron-rich
nuclei~\cite{savran2013,adrich2005,wieland2009,rossi2013}. Furthermore, strong M1 transitions are generated in deformed nuclei, giving rise to the scissors resonance (SR) at $E_{\gamma}=2-3$ MeV~\cite{heyde2010,krticka2004,guttormsen2012}. Finally, and very recently, a new feature has shown up in the $\gamma$-decay strength of light and medium-mass nuclei measured in charged-particle reactions: a very-low energy enhancement for $E_\gamma \lesssim 3-4$ MeV, see, e.g., Refs.~\cite{voinov2004,guttormsen2005,wiedekin2012,larsen2013}. This \textit{upbend} phenomenon was completely unforeseen,
and for long has had no satisfactory theoretical explanation.

Much progress has been made in the last few years towards a better understanding of the upbend. Its presence was shown in $^{95}$Mo in an Oslo-type experiment \cite{guttormsen2005} and confirmed in an independent experiment using a different technique ~\cite[and references therein]{wiedekin2012}.
Through angular-distribution measurements, it was demonstrated that the upbend is dominantly of dipole nature, excluding the possibility that it was caused by strong E2 transitions in the continuum~\cite{larsen2013}. On the theoretical side, the work of Litvinova and Belov~\cite{litvinova2013} suggested that the upbend is caused by thermal excitations in the continuum leading to enhanced low-energy E1 transitions. However, shell-model calculations~\cite{schwengner2013,brown2014,frauendorf2014} show very strong M1 transitions at low $\gamma$-ray energies.
Whether the upbend is of E1 or M1 character, or perhaps a mix of both, remains to be experimentally determined.

\begin{figure}[b]
\begin{center}
\includegraphics[clip,width=0.9\columnwidth]{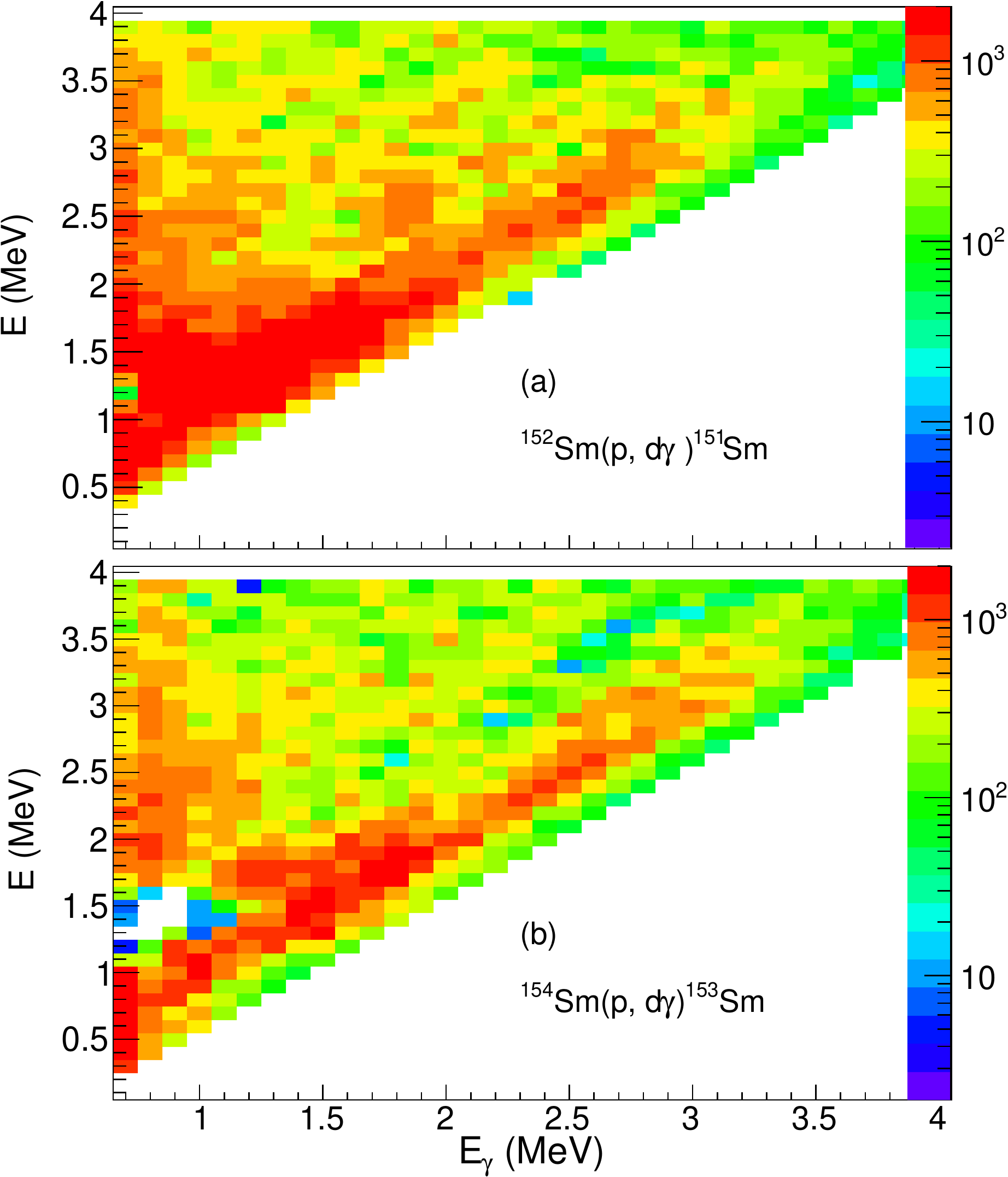}
\caption{(Color online) Primary $\gamma$-ray matrices of $^{151,153}$Sm.
For the Oslo method we use the following partition of the matrix:
$E_{\gamma}> 0.6$~MeV and $2.5 < E < 4.0$~MeV.}
\label{fig:matrices}
\end{center}
\end{figure}

\begin{figure*}[t]
\begin{center}
\includegraphics[clip,width=1.9\columnwidth]{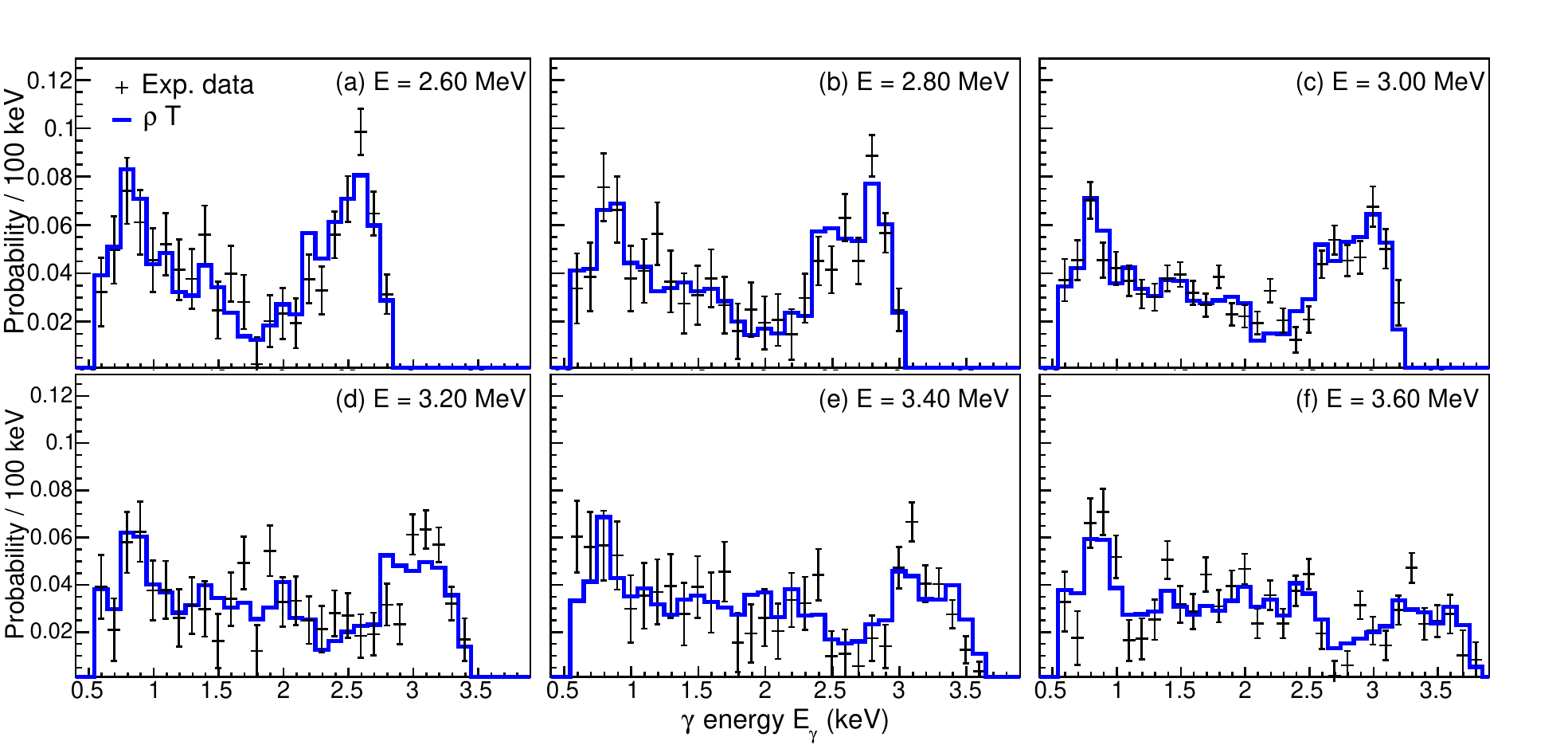}
\caption{(Color online) Primary $\gamma$-ray spectra (crosses) of $^{153}$Sm
from various initial excitation energies $E$ of 200 keV width. The spectra
are compared to the product $\rho(E-E_{\gamma}) {\mathcal{T}}(E_{\gamma})$ (blue histograms).}
\label{fig:does}
\end{center}
\end{figure*}

So far, as mentioned above, the upbend has been observed in light and medium-mass nuclei, with the heaviest
case being $^{138}$La~\cite{vincent2015}. Here, we present for the first time data that give evidence for
the upbend in the rare-earth region, more specifically in the $^{151,153}$Sm isotopes. The data were taken with Compton-suppressed Ge clover detectors, giving the opportunity to investigate the $\gamma$-decay strength below $\approx 1$~MeV, which has been the experimental limit in the Oslo-type experiments utilizing collimated NaI detectors. Moreover, as these Sm isotopes are deformed, we also see, for the first time, the presence of the SR \textit{and} the upbend in one and the same nucleus.
In the following sections, we will present the experimental details, the data analysis, and the results.

\section{Experimental procedure}
The experiment was performed at the Cyclotron Institute of Texas A\&M University, where two samarium targets, $^{152}$Sm and $^{154}$Sm, approximately 1 mg/cm$^2$ thick and 98(1)\% isotopically enriched were bombarded by a 1.2~nA of 25~MeV proton beam from the K-150 cyclotron. The reaction products were detected by the STARLiTeR setup \cite{Lesher,casperson2014} that consisted of a highly segmented $\Delta$E-E charged particle telescope and an array of six HPGe clover detectors with BGO Compton suppression for $\gamma$-ray detection. 

The telescope is comprised of two segmented silicon detectors, 140~$\mu$m ($\Delta$E) and 1000~$\mu$m (E) thick. Each of the detectors is a disk, 72~mm in diameter, with an 22~mm in diameter opening for the beam in the center. The disk is divided into 24 concentric 1~mm wide rings and into 8 segments in the angular direction. The $\Delta$E-E system was placed 18~mm behind the target, providing an angular coverage for particle detection of 30-58 degrees. The design of the telescope allowed for identification of the light ion charged particle reaction products (protons, deuterons and tritons) and an energy resolution of 130~keV FWHM for detected deuterons.

The clover $\gamma$-ray detectors were positioned approximately 13 cm from the target at 47, 90, and 133 degrees with respect to the incident beam axis. Using standard $\gamma$-ray calibration sources, an energy resolution of 2.6 keV and 3.5 keV FWHM was obtained at 122 keV and 963 keV, respectively. The absolute photopeak efficiency of the array was measured to be 4.8\% at 103 keV \cite{Humby}. Only the $\gamma$ rays coincident with a particle were recorded, which provided the data required to build the matrices for the Oslo method. The current study focused on two reactions: $^{152,154}$Sm(p,d$\gamma$)$^{151,153}$Sm.

\section{Extraction of the LD\lowercase{s} and $\gamma$SF\lowercase{s}}
The Oslo method determines simultaneously the functional form of the level density (LD) and $\gamma$-ray strength function ($\gamma$SF) without assuming any nuclear model. The first step is to sort the particle-$\gamma$ coincidences into
a matrix of initial excitation energy $E$ versus $\gamma$ energy. Then the matrix is unfolded~\cite{Gutt96} using the clover response function for each $E_\gamma$. The response functions were obtained from Geant4 \cite{geant} simulations of the STARLiTeR setup for $\gamma$ rays up to 10~MeV. In the next step, the primary $\gamma$ spectrum at $E$ is obtained by subtracting a weighted sum of unfolded spectra $U(E',E_{\gamma})$ at lower excitation energies $E'$:
\begin{equation}
P(E,E_{\gamma})=U(E,E_{\gamma}) - \sum_{E' < E}W(E,E')U(E',E_{\gamma}).
\label{eq:primary}
\end{equation}
The varying population cross section of the different excitation-energy bins is taken into account in a proper way. The weighting coefficients $W(E, E')$ are determined iteratively~\cite{Gutt87}: we first guess a $W$ distribution, then $P$ is calculated from Eq.~(\ref{eq:primary}) and replaces $W$ in the next iteration. After a few iterations, $W(E,E') \approx P(E,E_{\gamma})$ independent of the first $W$ trial function. This is exactly what is expected, namely that the primary $\gamma$-ray spectrum equals the weighting function. The technique is based on the assumption that the $\gamma$ distribution
is the same whether the levels were populated directly by the nuclear reaction or by $\gamma$ decay from higher-lying states.

 \begin{figure}[t]
 \begin{center}
 \includegraphics[clip,width=0.9\columnwidth]{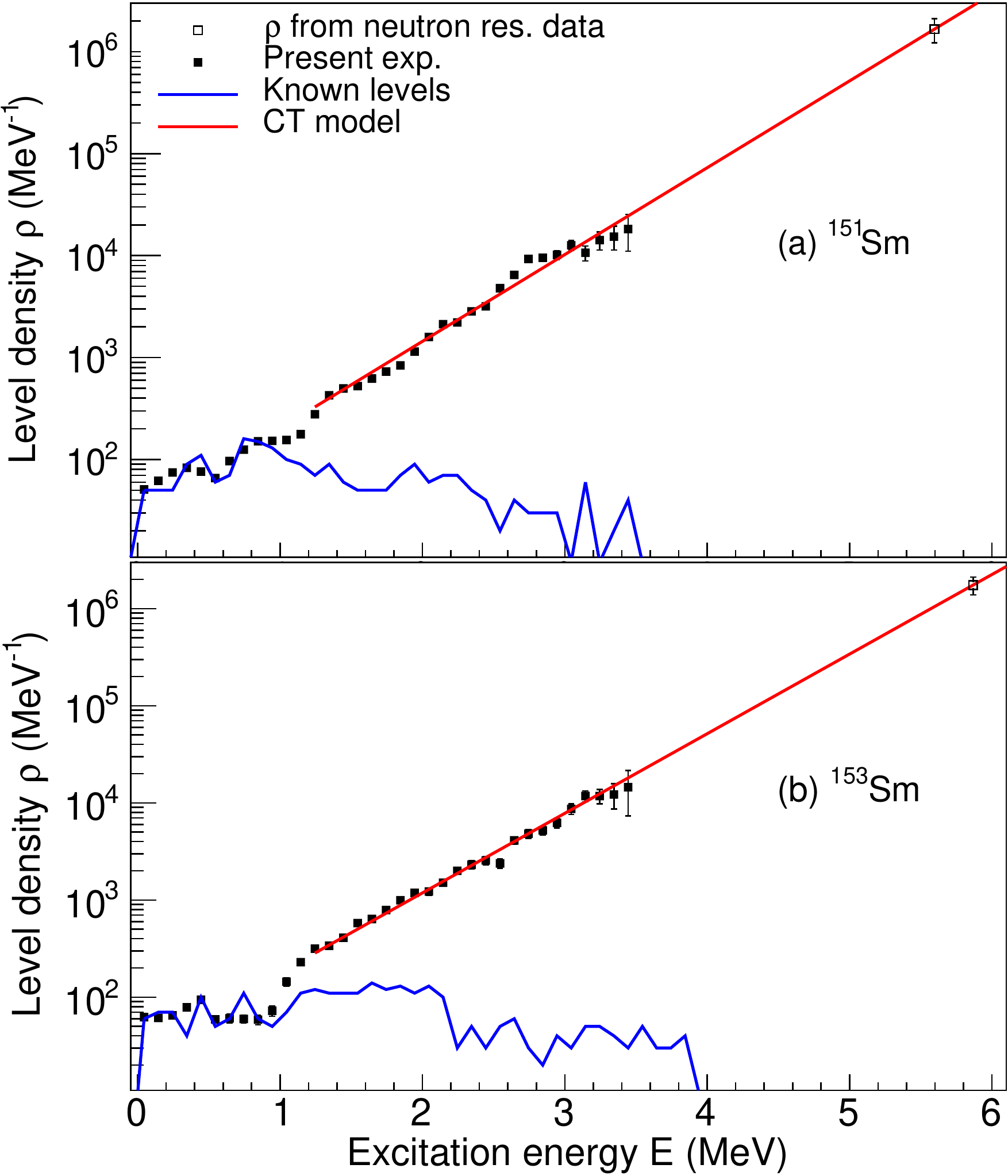}
 \caption{(Color online) Level densities for $^{151,153}$Sm. The experimental data (solid squares) are normalized to the LD of known discrete levels at low excitation energy $E$ (blue solid line) and to the LD extracted at the neutron separation energy $S_n$ (open square). The connection between $\rho(S_n)$ and our experimental data is performed with a constant-temperature LD formula (red line).}
 \label{fig:rho151_153}
 \end{center}
 \end{figure}

\begin{figure}[t]
\begin{center}
\includegraphics[clip,width=0.9\columnwidth]{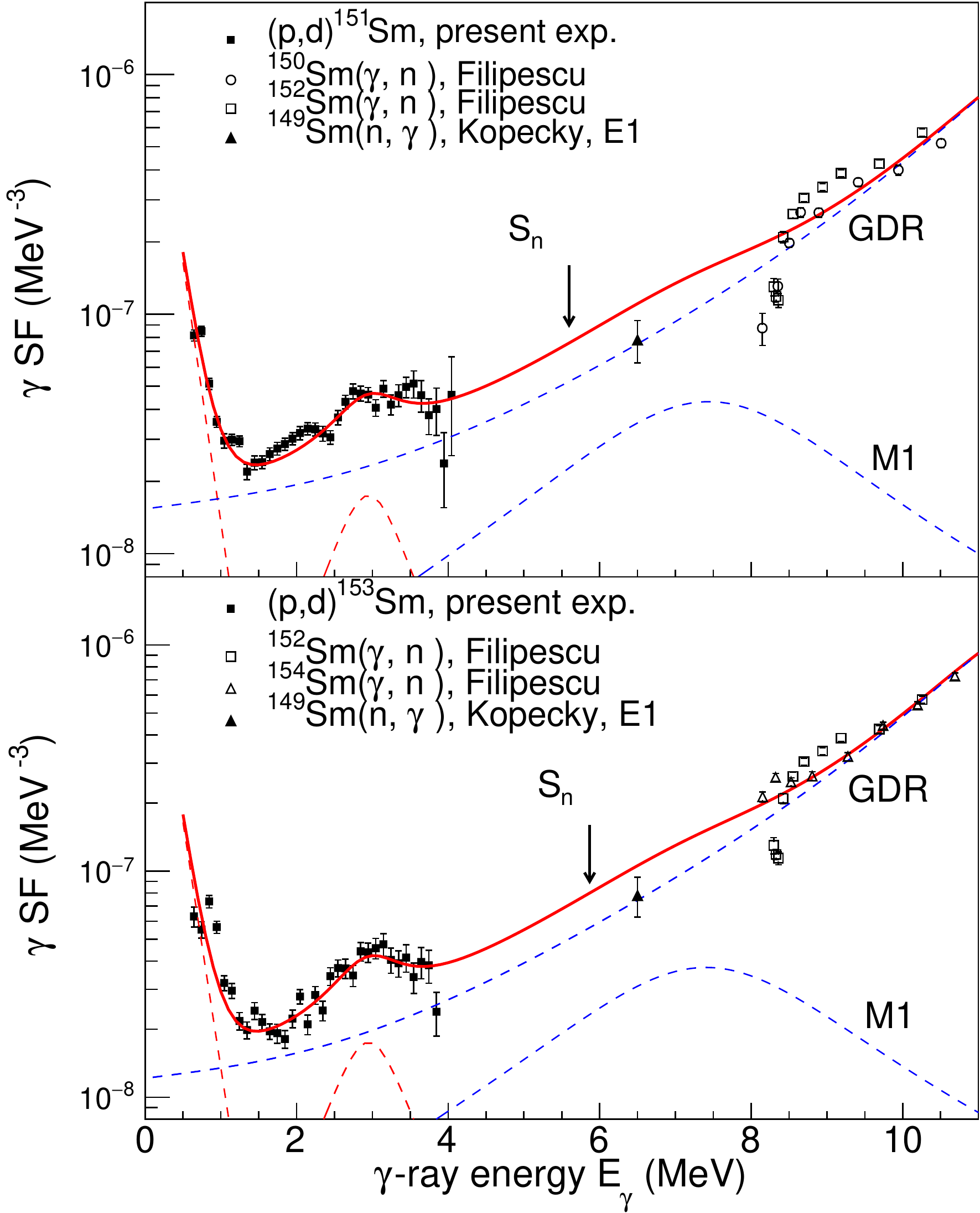}
\caption{(Color online) Experimental $\gamma$SF (solid squares) compared with summed (solid red line) contributions from: GDRs and spin-flip M1 strength (blue dashed curves) and the new observations of upbend and scissors (dashed red curves). The
($\gamma$, n) data (open circles and squares) are taken from Filipescu {\em et al.}~\cite{filipescu2014}. The filled triangle is based on the $^{149}$Sm(n,$\gamma$)$^{150}$Sm including only E1 strength, from RIPL-2~\cite{RIPL3}.}
\label{fig:GDR}
\end{center}
\end{figure}

    \begin{table}[]
    \caption{Parameters used to extract LD and $\gamma$SF.}
    \begin{tabular}{c|ccccc}
    \hline
    \multirow{2}{*}{Nucleus} &$S_n$ &  $\sigma(S_n)$&  $D_0$   &   $\rho(S_n)$    &  $\langle \Gamma_{\gamma}(S_n)\rangle$  \\
               &(MeV)     &               & (eV)     &(10$^6$MeV$^{-1}$)&            (meV)                        \\
    \hline
    $^{151}$Sm &5.597     & 6.15         & 46(8)   &    1.66(44)      &                60(5)                     \\
    $^{153}$Sm &5.868     & 6.31         & 46(3)   &    1.75(36)      &                60(5)                     \\
    \hline
    \end{tabular}
    \label{tab:parameters}
    \end{table}
    
The primary $\gamma$ matrices $P(E,E_{\gamma})$ for $^{151,153}$Sm are shown in Fig.~\ref{fig:matrices}, panels (a) and (b), respectively. According to the Brink-Axel hypothesis~\cite{brink,axel}, the $\gamma$-ray transmission coefficient ${\cal {T}}$ is approximately independent of excitation energy. Thus, the primary matrix may be factorized as follows:
\begin{equation}
P(E, E_{\gamma}) \propto   {\cal{T}}(E_{\gamma}) \rho (E -E_{\gamma}),\
\label{eqn:3}
\end{equation}
where $\rho (E -E_{\gamma})$ is the LD at the excitation energy after the first $\gamma$-ray has been emitted in the cascade.
This factorization allows the disentanglement of the LD and $\gamma$-ray transmission coefficient. Figure~\ref{fig:does} demonstrates that the product of the same ${\cal {T}}$ and $\rho$ functions describes the primary $\gamma$ spectra very well at six different excitation energies $E$. Thus, within the statistical errors the Brink-Axel hypothesis is valid and the factorization in Eq.~(\ref{eqn:3}) can be applied. This is in accordance with the recently found validity of the Brink-Axel hypothesis \cite{magne2016}.

\begin{table*}[]
\caption{Parameters for various resonances and the upbend, including the SR resonance strength (see text).}
\begin{tabular}{c|ccc|ccc|c|ccc|cc|ccc|c}
\hline
\multirow{3}{*}{Nucleus}&\multicolumn{7}{|c}{Giant dipole 1 and 2 resonances }&\multicolumn{3}{|c}{Spin-flip M1}&\multicolumn{2}{|c}{Upbend}&\multicolumn{4}{|c}{Scissors resonance}\\
\cline{2-17}
    &$\omega_{E1,1}$&$\sigma_{E1,1}$&$\Gamma_{E1,1}$&$\omega_{E1,2}$&$\sigma_{E1,2}$&$\Gamma_{E1,2}$&$T_f$&$\omega_{\rm M1}$&$\sigma_{\rm M1}$&$\Gamma_{\rm M1}$  &$C$                  &$\eta$         &$\omega_{\rm SR}$  &  $\sigma_{\rm SR}$&  $\Gamma_{\rm SR}$  &   $B_{\rm SR}$ \\
           &(MeV)          &      (mb)     &      (MeV)    &   (MeV)       &     (mb)      &    (MeV)      &(MeV)&          (MeV)    &      (mb)        &         (MeV)  &(MeV$^{-3}$)&(MeV$^{-1}$)   & (MeV)    & (mb)     &   (MeV)    & ($\mu_N^2$) \\
\hline
 $^{151}$Sm&  12.8         &      160      &       3.5     &     15.9      &      230      &     5.5       & 0.55&            7.7    &       3.8        &          4.0   &$20(10)10^{-7}$             &5.0(5)         &3.0(3)    & 0.6(2)   & 1.1(3)     &    7.8(34)  \\
 $^{153}$Sm&  12.1         &      140      &       2.9     &     16.0      &      232      &     5.2       & 0.45&            7.7    &       3.3        &          4.0   &$20(10)10^{-7}$             &5.0(10)        &3.0(2)    & 0.6(1)   & 1.1(2)     &    7.8(20)  \\
\hline
\end{tabular}
\\
\label{tab:GDR}
\end{table*}

\section{Normalization of the LD\lowercase{s} and $\gamma$SF\lowercase{s}}
In order to normalize the LD and $\gamma$SF we need to apply data from other experiments. For the normalization of the LD, we use two normalization points: ($i$) low excitation energy from the known level scheme~\cite{ENSDF} and ($ii$) high excitation energy from the density of neutron resonances following resonant ($n$, $\gamma$) capture at the neutron separation energy $S_n$. Here, the upper data point $\rho(S_n)$ is estimated from $\ell = 0$ neutron resonance spacings $D_0$ taken from RIPL-3~\cite{RIPL3} assuming the spin distribution of~\cite{GC}. The spin-cutoff parameter $\sigma$ was determined from the global systematic study of LD parameters by von Egidy and Bucurescu who use a rigid-body moment of inertia approach~\cite{egidy2}.

Figure \ref{fig:rho151_153} demonstrates how the LD is normalized to the anchor points at low and high excitation energies.
Above $E \approx 1.3$ MeV the LD follows roughly the constant-temperature LD formula~\cite{Ericson}
\begin{equation}
\rho_{\rm CT}(E)=\frac{1}{T_{\rm CT}}\exp{\frac{E-E_0}{T_{\rm CT}}},
\label{eq:ct}
\end{equation}
where $T_{\rm CT}$ is determined by the slope of $\ln \rho(E)$ and $E_0$ serves as a shift parameter, see the two red lines of Fig.~\ref{fig:rho151_153}. The fit parameters are ($T_{\rm CT},E_0)$ = (0.51, -1.37) MeV and (0.53, -1.41) MeV for $^{151,153}$Sm, respectively. A constant temperature behavior is the key characteristic of a first-order phase transition~\cite{luciano2014}.

The last step is to determine a scaling parameter for the transmission coefficient. The average, total radiative width $\langle \Gamma_{\gamma} \rangle$ at $S_n$ for initial spin $I$ and parity $\pi$ is given by~\cite{ko90}:
\begin{equation}
\langle\Gamma_\gamma\rangle=\frac{1}{2\pi\rho(S_n, I, \pi)} \sum_{I_f}\int_0^{S_n}{\mathrm{d}}E_{\gamma} B{\mathcal{T}}(E_{\gamma})\rho(S_n-E_{\gamma}, I_f),
\label{eq:norm}
\end{equation}
where the summation and integration run over all final levels with spin $I_f$ that are accessible by $E1$ or $M1$
transitions with energy $E_{\gamma}$. The scaling parameter $B$ for ${\cal {T}}(E_{\gamma})$ is adjusted to reproduce the experimental $\langle \Gamma_{\gamma} \rangle$. Details on the normalization procedure are given in Refs.~\cite{Schiller00,voin1}. The experimental data used for the normalizations are summarized in Table~\ref{tab:parameters}.

The dipole $\gamma$SF can be calculated from the transmission coefficient as ~\cite{RIPL3}:
\begin{equation}
 f (E_{\gamma}) =(1/ 2\pi)  ( {\cal {T}}(E_{\gamma}) / E_{\gamma}^3 ).
\end{equation} 
The data points of the $\gamma$SFs for $^{151,153}$Sm are displayed as solid squares in Fig.~\ref{fig:GDR}, panel (a) and (b), respectively. The figure also includes the $\gamma$SF derived from $^{150,152,154}$Sm($\gamma$, n) cross section data by Filipescu {\em et al.}~\cite{filipescu2014}. The transformation from photo-nuclear cross section $\sigma$ to $\gamma$SF is calculated from~\cite{RIPL3}:
\begin{equation}
f (E_{\gamma}) =(1 / 3\pi^2 \hbar^2c^2) (\sigma(E_{\gamma}) /  E_{\gamma}).
\end{equation} 

\begin{figure}[t]
\begin{center}
\includegraphics[clip,width=0.9\columnwidth]{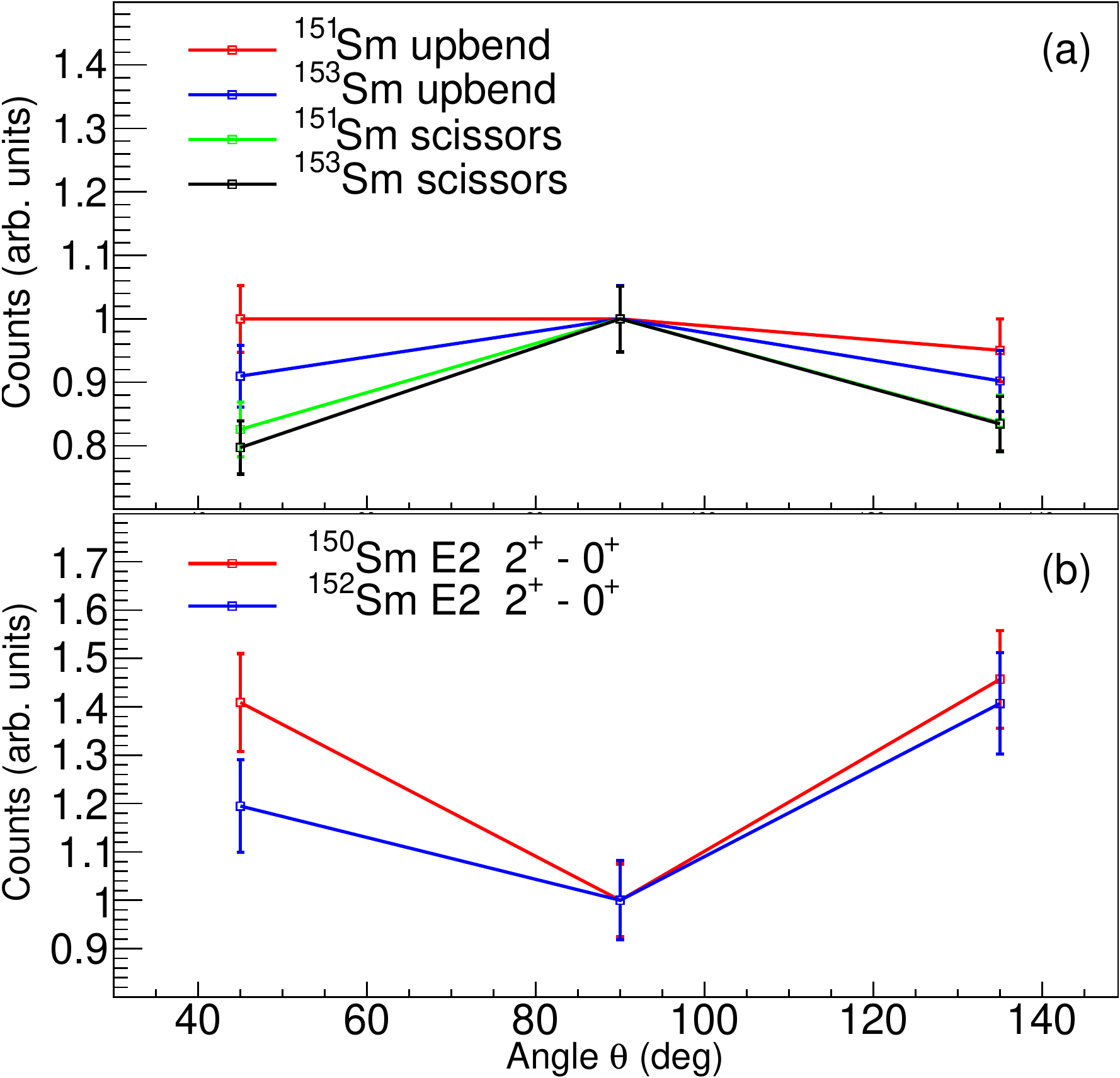}
\caption{(Color online) Gamma-ray angular distributions with respect to the beam direction of the upbend and SR structures in $(p,d)^{151,153}$Sm (a) compared with the
the E2 transitions of the directly populated $2^+$ states in $(d,t)^{150,152}$Sm (b). The upbend and SR data are taken from the
primary matrices at initial excitation energies $E=2.5-4.0$ MeV. The data are normalized to unity at 90$^\circ$.}
\label{fig:angular}
\end{center}
\end{figure}

Since our data cover $E_{\gamma} < 4$~MeV, we have to extrapolate the ($\gamma$, n) data in order to match our data. For the double-humped giant electric dipole resonance (GDR) we fit the data with two generalized Lorentzians (GLO) as defined in RIPL-3~\cite{RIPL3}. The M1 spin-flip resonance with a Lorentzian shape is also taken from RIPL-3~\cite{RIPL3},
but with adjusted strength ($\sigma_{\rm M1}$) in order to obtain reasonable fit with the high-energy part of the present data points. The summed GDRs and the M1 spin-flip $\gamma$SFs are shown as blue dashed curves in Fig.~\ref{fig:GDR}.
The three sets of resonance parameters are listed in Table~\ref{tab:GDR}.

\section{Enhancement in the $\gamma$SF}
The measured $\gamma$SFs of Fig.~\ref{fig:GDR} show two pronounced structures: a low-energy enhancement and a bump centered at $E_{\gamma}\approx 3$~MeV. Figure \ref{fig:angular} indicates that the angular distributions of these structures are of dipole type, contrary to the E2 distributions for the $2^{+} \rightarrow 0^{+}$ transitions in the neighbouring $^{150,152}$Sm isotopes. The multipolarity, energy position and strength of these structures support their interpretations as the {\em upbend} and the {\em scissors resonance} (SR). To our knowledge, the SR is the only known candidate for a soft collective mode at ${\gamma}$ energies around 3 MeV. This is supported by nuclear resonance fluorescence experiments, which demonstrate strong M1 transitions at these $\gamma$-ray energies \cite{enders2005}.

Applying the GDR and M1 spin-flip parameterization as described above, we can model the upbend and the SR. Based on empirical data for lighter nuclei and shell model calculations we may describe the upbend by:
\begin{equation}
\label{eqn:upbend}
f_{\rm upbend}(E_{\gamma}) = C\exp(-\eta E_{\gamma}).
\end{equation}
For the SR we use the Lorentzian shape. The results for the two low-energy structures are shown as dashed red curves in Fig.~\ref{fig:GDR} with parameters listed in Table~\ref{tab:GDR}. In the last column, the strength of the SR is calculated as
$B=(9\hbar c/32 \pi ^2)( \sigma \Gamma /  \omega )$ giving values comparable with the results of other rare-earth nuclei in
the quasicontinuum~\cite{yb2001a,yb2001b,dy2003,yb2004,dy2010}.

Previously, it was shown for the actinides~\cite{tornyi2014,guttormsen2014} that the energy centroid and strength are well described by the sum-rule approach of J.~Enders {\em et al}.~\cite{enders2005}.  Here, the inversely and linearly energy-weighted sum rules, $S_{+1}$ and  $S_{-1}$, give $ \omega_{\rm SR}=\sqrt{ S_{+1}/ S_{-1}}$ and $B_{\rm SR}=\sqrt{ S_{+1}S_{-1}}$. Assuming a rigid moment of inertia and a deformation of $\delta = 0.33$ for $^{151,153}$Sm, we obtain $\omega_{\rm SR}=3.0$~MeV and $B_{\rm SR}=7.3 \mu{_N}^2$ in good agreement with the experimental findings. These sum rules are also consistent with the results of other rare-earth nuclei in
the quasicontinuum~\cite{yb2001a,yb2001b,dy2003,yb2004,dy2010}.

The upbend and scissors structures are clearly separated in $\gamma$ energy, indicating that they originate from different mechanisms. It is possible that the upbend has a similar origin as the shears bands mechanism~\cite{schwengner2013}, but not only for high spins as it is for the magnetic rotation. It could also be that it is present for all aligned high-$\ell$ orbitals, i.e. proton-proton, neutron-neutron or proton-neutron configurations, independent of their particle-hole nature~\cite{brown2014}. In the latter case, the upbend would be expected throughout the whole chart of nuclei, and both the upbend and the SR would stem from 0$\hbar \omega$ transitions between orbitals within the same shell.

The SR has components of large transitions between magnetic sub-states differing with one unit of angular momentum. More specifically,  the transitions correspond to $\Omega  \rightarrow \Omega \pm 1$ transitions with similar spherical $j$-components in the Nilsson scheme. The energy splitting between these Nilsson orbitals is proportional to the
nuclear deformation and is the reason for the higher and well-separated $\gamma$-energy centroid of $2-3$~MeV~\cite{guttormsen1984}. For transitional nuclei with low deformation, we foresee an exciting situation  where the upbend and the SR merge together in a new type of structure.

The present $\gamma$SF includes both the upbend and the GDR tail being responsible for a minimum strength at $E_{\gamma}\approx 1.2$~MeV. The corresponding $\gamma$SF minima for $^{56}$Fe, $^{92-98}$Mo and $^{138}$La are approximately 4, 3 and 2~MeV, respectively~\cite{larsen2013,guttormsen2005,vincent2015}. From these systematics, the minima are expected to
disappear for nuclei with mass numbers $A > 200$. However, the low-energy enhanced M1 transitions are probably still present, but the strength is overwhelmed by E1 transitions from the relatively strong tail of the GDR. A great challenge  would be to design experiments to reveal the M1 part of the low-energy $\gamma$s for the heavier nuclei.

\section{(\lowercase{n},$\gamma$) reaction rates}
To investigate the impact of the upbend and the SR on astrophysical $(n,\gamma)$ reaction rates, we have performed calculations with the nuclear reaction code TALYS~\cite{TALYS}. The most important ingredients into these calculations are the nuclear level density, the $\gamma$-ray strength function, and the neutron optical-model potential (n-OMP), as well as the masses and deformations for the very exotic, neutron-rich Sm isotopes. For consistency, we have chosen input models for the masses, the level density and the E1 strength from one and the same framework, i.e. the Skyrme-Hartree-Fock-Bogoliubov (HFB) plus combinatorial model for the level density~\cite{goriely2008}, the Skyrme-HFB approach for the nuclear masses and deformations~\cite{goriely2009}, and the Skyrme-HFB plus quasiparticle random-phase approximation for the microscopic E1 strength function~\cite{goriely2003,goriely2004}. For the n-OMP we have used the global parameterization of Ref.~\cite{koning03}. For the M1 part of the strength, the standard treatment of the M1 spin-flip transitions is applied (see the TALYS manual~\cite{TALYS}), and we have added the scissors resonance with centroid and summed strength according to the sum rules described previously, assuming a width of 1.1 MeV. Moreover, we have assumed that the upbend can be parameterized as  in Eq.~(\ref{eqn:upbend}) for all Sm nuclei and using the same parameters as for $^{151,153}$Sm, i.e. $C = 20 \cdot 10^{-7}$ MeV$^{-3}$ and $\eta = 5.0$ MeV$^{-1}$.

\begin{figure}[t]
\begin{center}
\includegraphics[clip,width=0.9\columnwidth]{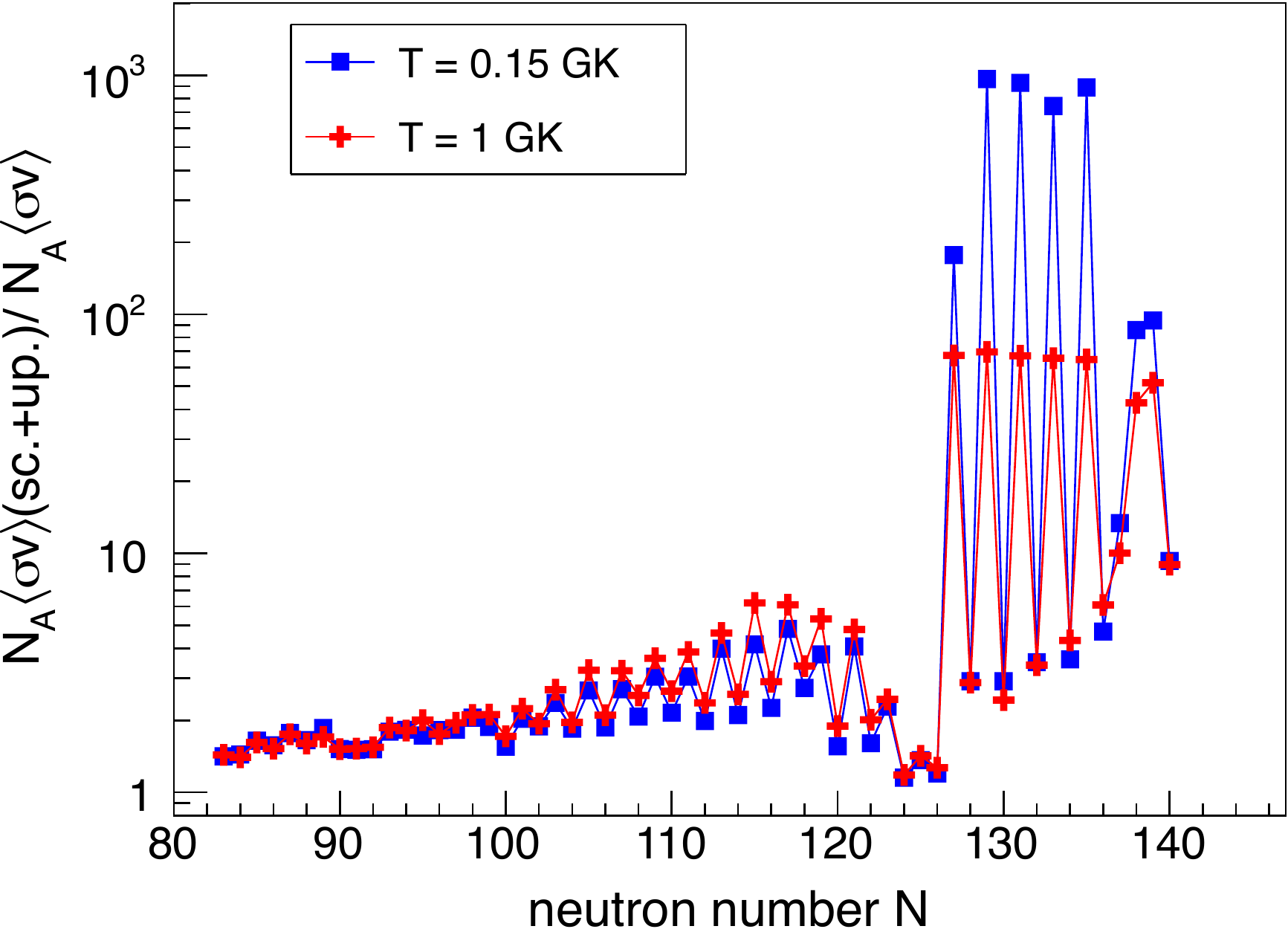}
\caption{(Color online) Ratios of Maxwellian-averaged (n,$\gamma$ ) reaction rates at $T = 0.15$ and $1.0$ GK  for the Sm isotopic chains up to the neutron drip line, see text.}
\label{fig:astro}
\end{center}
\end{figure}

The resulting ratios of the $(n,\gamma)$ reaction rates including the SR and upbend divided by the reaction rates without these M1 components are shown in Fig.~\ref{fig:astro} for two different temperatures of an (unknown) r-process site, $T = 0.15$ and $1.0$ GK. The increase in reaction rate is striking, in particular for the very neutron-rich nuclei across the $N=126$ shell gap and for the “cold” stellar temperature of 0.15 GK, where an enhancement of up to 3 orders of magnitude is seen. Also for the cases where $N<126$, a significant increase is observed. Hence, we conclude that there is, potentially, a non-negligible effect on the astrophysical reaction rates, provided a similar M1 strength in neutron-rich Sm isotopes as for $^{151,153}$Sm.

\section{Summary}
In conclusion, the level densities and $\gamma$-ray strength functions of $^{151,153}$Sm have been determined for the first time using the Oslo method with clover detectors. The Compton suppression of the $\gamma$ detectors allowed for exploration of the low energy range of the $\gamma$SF not accessible for other types of experiments utilizing the Oslo method. For the first time, the low-energy $\gamma$ enhancement has been observed for rare-earth nuclei. The upbend coexists with the scissors resonance indicating that the two structures originate from different mechanisms and are not mutually exclusive. The observed low-energy part of the $\gamma$SF may play a major role for the $(n,\gamma)$ cross sections for the very neutron-rich Sm isotopes involved in the r-process nucleosynthesis.

\acknowledgements
The authors would like to thank ORNL for providing one of the clover detectors for the STARLiTeR setup.
This work was supported by the US Department of Energy under grant numbers DE-NA0001801, DE-FG02-05ER41379 and DE-AC52-07NA27344, the Research Council of Norway (NFR), and the National Science Foundation under grant number PHY-1430152 (JINA-CEE). A.C.L. acknowledges financial support from the Research Council of Norway, grant no. 205528, and funding from the ERC-STG-2014 under grant agreement no. 637686.


\begin{thebibliography}{99}

\bibitem{heyde2011}K.~Heyde and J.L.~Wood, Rev.~Mod.~Phys.~{\bf 83}, 1467 (2011), and references therein.
\bibitem{dietrich1988}S.S.~Dietrich and B.L.~Berman, At. Data Nucl. Data Tables {\bf 38}, 199 (1988).
\bibitem{savran2013} D.~Savran, {\em et al.}, Prog. Part. Nucl. Phys. {\bf 70}, 210 (2013).
\bibitem{adrich2005}P.~Adrich, {\em et al.}, Phys.\ Rev.\ Lett. \bf 95\rm, 132501 (2005).
\bibitem{wieland2009}O.~Wieland, {\em et al.}, Phys.\ Rev.\ Lett. \bf 102\rm, 092502 (2009).

\bibitem{rossi2013}D.M.~Rossi, {\em et al.}, Phys.\ Rev.\ Lett. \bf 111\rm, 242503 (2013).
\bibitem{heyde2010}K.~Heyde, P.~von Neumann-Cosel, A.~Richter, Rev.~Mod.~Phys.~{\bf 82}, 2365 (2010), and references therein.
\bibitem{krticka2004}M.~Krti\u{c}ka, {\em et al.}, Phys.\ Rev.\ Lett. \bf 92\rm, 172501 (2004).
\bibitem{guttormsen2012}M.~Guttormsen, {\em et al.}, Phys.\ Rev.\ Lett. \bf 109\rm, 162503 (2012).
\bibitem{voinov2004}A.~Voinov, {\em et al.}, Phys.\ Rev.\ Lett. \bf 93\rm, 142504 (2004).

\bibitem{guttormsen2005}M.~Guttormsen, {\em et al.}, Phys.\ Rev.\ C \bf 71\rm, 044307 (2005).
\bibitem{wiedekin2012}M.~Wiedeking, {\em et al.}, Phys.\ Rev.\ Lett. \bf 108\rm, 162503 (2012).
\bibitem{larsen2013}A.C.~Larsen, {\em et al.}, Phys.\ Rev.\ Lett. \bf 111\rm, 242504 (2013).
\bibitem{litvinova2013}E.~Litvinova and N.~Belov, Phys.\ Rev.\ C \bf 88\rm, 031302(R) (2013).
\bibitem{schwengner2013}R.~Schwengner, S.~Frauendorf, and A.C.~Larsen, Phys.\ Rev.\ Lett. \bf 111\rm, 232504 (2013).

\bibitem{brown2014}B.A.~Brown and A.C.~Larsen, Phys.\ Rev.\ Lett. \bf 113\rm, 252502 (2014).
\bibitem{frauendorf2014}S. Frauendorf, R. Schwengner and K. Wimmer, AIP Conf. Proc. \bf 1619\rm, 81 (2014).
\bibitem{vincent2015}B.V.~Kheswa, {\em et al.}, Phys.\ Lett. \ B {\bf 744}, 268 (2015).

\bibitem{Lesher} S.~Lesher, {\em et al.}, Nucl.~Instrum.~Methods Phys.~Res.~A {\bf 621}, 286 (2010).
\bibitem{casperson2014} R. J. Casperson, {\em et al.}, Phys.\ Rev.\ C \bf 90\rm, 034601 (2014).
\bibitem{Humby} P.~Humby, {\em et al.}, Phys.\ Rev.\ C {\bf 91}, 024322, (2015).

\bibitem{Gutt96} M.~Guttormsen, T.S.~Tveter, L.~Bergholt, F.~Ingebretsen, and J.~Rekstad,Nucl.\ Instrum.\ Methods Phys.\ Res.\ A \bf 374\rm, 371 (1996).

\bibitem{geant} S. Agostinelli, {\em et al.}, Nucl.\ Instrum.\ Methods Phys.\ Res.\ A {\bf 506}, 250 (2003).

\bibitem{Gutt87}M.~Guttormsen, T.~Rams{\o}y, and J.~Rekstad, Nucl.\ Instrum.\ Methods Phys.\ Res.\ A \bf 255\rm, 518 (1987).
\bibitem{brink} D.M.~Brink, Ph.D.~thesis, Oxford University (1955).
\bibitem{axel}P.~Axel, Phys. Rev. {\bf 126}, 671 (1962).
\bibitem{magne2016} M. Guttormsen,  A. C. Larsen,  A. G\"orgen,  T. Renstr{\o}m,  S. Siem,  T. G. Tornyi,  and G. M. Tveten, Phys. Rev. Lett. {\bf 116} (2016) 012502. 

\bibitem{ENSDF}Data extracted using the NNDC On-Line Data Service from the ENSDF database.
\bibitem{RIPL3}R.~Capote, {\em et al.}, Reference Input Parameter Library, RIPL-2 and RIPL-3, available online at {\it http://www-nds.iaea.org/RIPL-3/}
\bibitem{GC} A.~Gilbert and A.G.W.~Cameron, Can. J. Phys. {\bf 43}, 1446 (1965).
\bibitem{egidy2}T.~von Egidy and D.~Bucurescu, Phys.\ Rev.\ C \bf 72\rm, 044311 (2005); Phys.\ Rev.\ C \bf 73\rm, 049901(E) (2006).
\bibitem{Ericson}T.~Ericson, Nucl. Phys. A {\bf 11}, 481 (1959).
\bibitem{luciano2014}L.G.~Moretto, A.C. Larsen, F. Giacoppo, M. Guttormsen, S. Siem, and A.V. Voinov, arXiv:1406.2642 [nucl-th] (2014).

\bibitem{ko90}J.~Kopecky and M.~Uhl, Phys.~Rev.~C {\bf 41} 1941 (1990).
\bibitem{Schiller00}A.~Schiller, {\em et al.}, Instrum.~Methods Phys.~Res.~A {\bf 447}, 498 (2000).
\bibitem{voin1}A.~Voinov, M. Guttormsen, E. Melby, J. Rekstad, A. Schiller, and S. Siem, Phys.\ Rev.\ C \bf 63\rm, 044313 (2001).
\bibitem{filipescu2014}D.M.~Filipescu, {\em et al.}, Phys.~Rev.~C {\bf 90}, 064616 (2014).

\bibitem{guttormsen2014}M.~Guttormsen, {\em et al.}, Phys.~Rev.~C {\bf 89}, 014302 (2014).
\bibitem{tornyi2014}T.G.~Tornyi, {\em et al.}, Phys.~Rev.~C {\bf 89}, 044323 (2014).

\bibitem{enders2005}J. Enders, P. von Neumann-Cosel, C. Rangacharyulu, and A. Richter, Phys.~Rev.~C {\bf 71}, 014306 (2005).

\bibitem{yb2001a}A.~Schiller, {\em et al.}, Phys.\ Rev.\ C \bf 63\rm, 021306(R) (2001).
\bibitem{yb2001b}A.~Voinov, {\em et al.}, Phys.\ Rev.\ C \bf 63\rm, 044313 (2001).
\bibitem{dy2003}M.~Guttormsen, {\em et al.}, Phys.~Rev.~C {\bf 68}, 064306 (2003).
\bibitem{dy2010}H.T.~Nyhus, {\em et al.}, Phys.\ Rev.\ C \bf 81\rm, 024325 (2010).
\bibitem{yb2004}U.~Agvaanluvsan, {\em et al.}, Phys.\ Rev.\ C \bf 70\rm, 054611 (2004).

\bibitem{guttormsen1984}M.~Guttormsen, J.~Rekstad, A.~Henriquez, F.~Ingebretsen, and T.F.~Thorsteinsen, Phys.~Rev.~Lett.~{\bf 52}, 102 (1984).

\bibitem{TALYS} A.~J.~Koning, S.~Hilaire and M.~C.~Duijvestijn, “TALYS-1.6”,
\textit{Proceedings of the International Conference on Nuclear Data for Science and Technology},
April 22-27, 2007, Nice, France, editors O.~Bersillon, F.~Gunsing, E.~Bauge, R.~Jacqmin, and S.~Leray,
EDP Sciences, 2008, p. 211-214.
\bibitem{goriely2008} S. Goriely, S. Hilaire, and A. J. Koning, Phys. Rev. C \textbf{78}, 064307 (2008).
\bibitem{goriely2009} S. Goriely, N. Chamel, and J. M. Pearson, Phys. Rev. Lett \textbf{102}, 152503 (2009).
\bibitem{goriely2003} S. Goriely, M. Samyn, M. Bender and J. M. Pearson, Phys. Rev. C \textbf{68} 054325 (2003).
\bibitem{goriely2004} S. Goriely, E. Khan, and M. Samyn, Nucl. Phys. \textbf{A739}, 331 (2004).
\bibitem{koning03} A. J. Koning and J.-P. Delaroche, Nucl. Phys. \textbf{A713}, 231 (2003).
\end{thebibliography}
\end{document}